\documentclass[useAMS,usenatbib]{mn2e}
\usepackage{graphicx}
\usepackage{latexsym,amsmath,amssymb}
\usepackage{natbib}
\usepackage{rotating}
\bibpunct{(}{)}{;}{a}{}{,}
\def \degmark{^\circ}
\def \nh {N${\rm _H}$}

\def \hcm {\hbox {\ifmmode $ atom cm$^{-2}\else atom cm$^{-2}$\fi}}

\def \ecut {\hbox {$E{\rm _{cut}}$}}
\def \chisq {$\chi ^{2}$}

\def\approxgt{\mathrel{\hbox{\rlap{\lower.55ex \hbox {$\sim$}}
        \kern-.3em \raise.4ex \hbox{$>$}}}}
\def\approxlt{\mathrel{\hbox{\rlap{\lower.55ex \hbox {$\sim$}}
        \kern-.3em \raise.4ex \hbox{$<$}}}}

\def \XMM {$XMM$-$Newton$\ }

\def \thirteen {XB\,1323$-$619}

\def \nh {$N{\rm _H}$}

\title[Frequency Resolved Spectroscopy of \thirteen]
{Frequency Resolved Spectroscopy of \thirteen\ Using \emph{XMM-Newton} data: 
Detection of a Reflection Region in the Disk} 
\author[\c{S}. Balman]{\c{S}. Balman$^{1, 2}$\thanks{E-mail:\,solen@astroa.physics.metu.edu.tr}
\\
$^{1}$Department of Physics, Middle East Technical University,
In\"on\"u Bulvar{\i}, Ankara, Turkey\\
$^{2}$Department of Physics, University of Warwick, Coventry CV4 7AL, UK
}
\begin{document}

\pagerange{\pageref{firstpage}--\pageref{lastpage}} \pubyear{2009}

\maketitle

\label{firstpage}

\begin{abstract}

We present the frequency resolved energy spectra (FRS) of the low-mass
X-ray binary dipper \thirteen\ during persistent emission 
in four different frequency bands
using an archival \XMM observation. FRS method helps to
probe the inner zones of an accretion disk.
\thirteen\ is an atoll source  and a type-I burster.
We find that the
FRS is well described by a single blackbody component with kT
in a range 1.0-1.4 keV responsible for the source variability in 
the frequency ranges 0.002-0.04 Hz, and 0.07-0.3 Hz. 
We attribute this component to the accretion disk and possibly emission from 
an existing boundary layer supported by radiation pressure. The 
appearance of the blackbody component in the
lower frequency ranges and disappearance towards the higher frequencies
suggests that it may also be a disk-blackbody emission.
We detect a different form of FRS for the higher frequency
ranges 0.9-6 Hz and 8-30 Hz  which is modeled best with a power-law 
and a Gaussian emission line at 6.4$^{+0.2}_{-0.3}$ keV with an equivalent width 
of 1.6$^{+0.4}_{-1.2}$ keV and 1.3$^{+0.7}_{-0.9}$ keV for the two frequency 
ranges, respectively. 
This iron fluorescence line detected in the higher frequency ranges of spectra shows the existence of
reflection in this system within the inner disk regions. 
The conventional spectrum of the source
also shows a weak broad emission line around 6.6 keV (Boirin et al. 2005). 
In addition, we find that 
the 0.9-6 Hz frequency band shows two QPO peaks at 1.4$^{+1.0}_{-0.2}$ Hz
and 2.8$^{+0.2}_{-0.2}$  Hz at about 2.8-3.1 $\sigma$ confidence level. These are
consistent with the previously detected $\sim$ 1 Hz QPO
from this source (Jonker et al. 1999). We believe they relate to the
reflection phenomenon. 
The emission from the reflection region, being a variable spectral component 
in this system, originates
from the inner regions of the disk with a maximum size of 4.7$\times 10^9$ cm
and a minimum size of 1.6$\times 10^8$ cm calculated using
 light travel time considerations and our frequency resolved spectra.
   
\end{abstract}


\section{Introduction}

\thirteen\ is an X-ray burster that shows varying intensity dips
that repeat with the orbital period. The source is discovered by {\it Uhuru}
and {\it Ariel V} (Forman et al. 1978; Warwick et al. 1981) and the dips
and bursts are recovered with EXOSAT (van der Klis et al. 1985;
Parmar et al. 1989). The typical dips last about 30\%
of the orbital cycle, and the 0.3-12.0 keV intensity varies creating
sharp deep features in time 
with about $\sim$70\% decrease in the persistent emission count rate. 
The source is viewed at $i = $
60--80$\degmark$ (Frank et al. 1987). The dip recurrence interval is
not precisely known, with the best measurement of 2.94(2)~hr
(Balucinska-Church et al. 1999). The source
is found to have  persistent $\sim$1~Hz quasi-periodic oscillations (QPOs) (Jonker et al. 1999). 

A $BeppoSAX$ observation of \thirteen\ between 1.0--150~keV
 reveals persistent emission of the source  modeled by a
cutoff power-law with $\alpha$ = $1.48 \pm 0.01$ and \ecut\ =
$44.1 \, ^{+5.1} _{-4.4}$~keV together with a blackbody with $kT$
= $1.77 \pm 0.25$~keV (Balucinska-Church et al. 1999). A more recent
$INTEGRAL$ time-averaged 4.0-200.0~keV spectrum (combined JEM-X and 
ISGRI spectra) of \thirteen\ 
can be best fitted using a
blackbody emission of $kT$ = $1.7\pm 0.3$~keV and a
Comptonized plasma emission model (CompTT, Titarchuck 1994) of  
$kT$ = $196.4\pm 28.5$ keV 
with a seed in photon temperature of $0.4\pm 0.2$~keV, 
and a Compton optical depth to scattering
of $\tau$ = 0.002-0.007 (Balman  2009). An $RXTE$ observation below 20 keV
reveals a 
cutoff power-law with $\alpha$ = $1.23 \pm 0.07$ and a blackbody with $kT$
= $1.36 \pm 0.06$~keV for the persistent emission (Barnard et al. 2001). 
A recent $Suzaku$ observation
between 0.8 and 70 keV has a non-dip spectrum consistent with a blackbody model of
$kT$ = $1.35 \pm 0.36$~keV and a power-law with an $\alpha$ = $1.67 \pm 0.1$
(Balucinska-Church et al. 2009). 
Boirin et al. (2005) detect Fe XXV and Fe XXVI absorption features from this source
using the same archival observation analyzed in this paper. 
They find changes in the properties of
the Fe XXV and Fe XXVI absorption features from persistent
to dipping intervals indicating the presence of
less-ionized material in the line of sight during dips 
(ionization parameter $\xi$ decreases from log($\xi$) of $3.9\pm 0.1$
to log($\xi$) of $3.13\pm 0.07$).  It is suggested that
absorption lines in dipping low-mass X-ray binaries are a 
result of a circularly symmetric photoionized
region on the outer disk and as the source is viewed from the absorber at all times, 
the variations in the ionization parameter of this region causes the different
levels of dipping in these systems (see also Diaz-Trigo et al. 2006, 2009).
Church et al. (2005) has studied the same data set as Boirin et al. (2005) 
using absorption line ratios of Fe XXV/Fe XXVI  
and curve of growth analysis leading to the conclusion that a 
collisionally ionized plasma with kT=31 keV 
can reproduce
the absorption  lines in the \XMM Spectra. As they find that this temperature is close to the accretion disk corona
(ADC) temperature (they have calculated using the $BeppoSax$ spectrum), they propose that the absorption
lines in dipping low-mass X-ray binaries are produced in the ADC and there is no need to invoke
a separate location where a highly ionized absorber exists.

In this paper, we aim at deriving frequency resolved spectra of \thirteen\ in an attempt to
reveal different emission regions and  absorbing regions in the system extracting the 
variable component of the emission spectrum as opposed to the non-variable component by the described 
method in the paper. We expect that this method should, also, reveal the location of 
the emitting and absorbing regions 
in similar (dipping) LMXBs.

\section{Data and the Observation}

The \XMM Observatory (Jansen et al. 2001) has three 1500 cm$^2$ X-ray telescopes
each with an European Photon Imaging Camera (EPIC) at the focus; two of which have MOS CCDs
(Turner et al. 2001) and the last one uses pn CCDs (Str\"uder et al. 2001) for data recording.
Also, there are two Reflection Grating Spectrometers (RGS, den Herder et al. 2001) located
behind two of the telescopes. We use an archival \XMM observation of \thirteen\ 
with a duration of 50 ksec between
2003 January 29 09:05 UTC and 2003 January 29 28:58 UTC.
A thin optical blocking filter was used with the EPIC Cameras. 
EPIC pn and MOS1 were operated  in the timing  mode whereas MOS2 was operated in the full window imaging mode. 
For the FRS calculations in our work, we used only the timing mode EPIC pn data which provides a 
time resolution of 30 $\mu$sec together with the highest sensitivity (i.e., count rate) and energy 
resolution among the
EPIC CCDs. 
The EPIC pn timing mode is a mitigation for pile-up in high count rate sources
and allow for pile-up free data for rates below 1500 c s$^{-1}$ compressing the data 
to a single one-dimensional row of length 4$^{\prime}$.4.  
We analyzed the pipeline-processed EPIC pn timing mode data using Science Analysis Software ({\tt SAS}) 
version 8.0.5. Data for analysis
(single- and double-pixel events, i.e., patterns 0--4 with Flag=0 option) were extracted from a rectangular 
region of 87$^{\prime\prime}$ wide column for the source and 
the background events were extracted whenever necessary from a
source free zone normalized to the source extraction area (ie.e., same rectangular area). We calculated
a conventional source spectrum using the SAS task {\tt ESPECGET} which automatically creates 
a source spectrum, background spectrum,
the ancillary response file and the energy response matrix. These files were later utilized to calculate the 
response files
for the frequency resolved spectra as the spectra were derived with the method explained in the next section.
In general, background subtraction was applied for the FRS since EPIC pn background contribution 
at frequencies 0.002-300 Hz would be negligible. 
In the analysis, only persistent emission (as opposed to dipping parts of the light curve) 
were used to calculate light curves for the FRS analysis. In order to filter
the persistent emission user-defined good time intervals were created using thresholds on count 
rates resulting in an effective exposure  of about 40 ksec. 
Data were also checked for flaring episodes (of the background)
 and type-I bursts of the source and all such occurrences were 
cleaned with the above procedure. We obtained an average count rate of 
27.3$\pm$0.02 for the source with a maximum 
variation of 25$\%$ for the persistent emission.
This was later used to calibrate systematic errors of 13$\%$ on the FRS spectral bins.

\section{Analysis and Results}


\subsection{Frequency Resolved Spectra}

In this work, the frequency resolved energy spectra were calculated following the prescription
of Revnivtsev et al. (1999) and Gilfanov et al. (2003) where they applied this method to
RXTE data of LMXBs. One assumes that the inverse Fourier Transform, the light curve,
is a function of an energy dependent term and a time dependent term and 
the light curves at different energies are related by linear transformations.
One allows for two components in the source emission spectrum,
a constant non-variable part of the source emission and flux variations of the second variable
component. Spectrum of the variable component is the frequency resolved spectrum, FRS. 
In order to calculate the FRS, we created 
power spectra (PSD--Power Spectral Density) in
 a certain range of energy channels (100-4095) of the \XMM EPIC pn data with 
a pre-determined binning of 50, 100
or 200 energy channels in the given range for 320-16 seconds time segments for each PSD. 
In each set of four frequency ranges, 
spectra were produced with the same 
channel binning and the duration for the PSD segments. 
Fast Fourier Transforms (FFT)
were computed to derive the PSD adopting the normalization of Miyamoto et al. (1991) 
$$ P_j=2|A_j|^2/N_{ph}C\ \ \ \ \ \  
A_j=\sum x_n e^{i w_n t_n} $$. In this prescription, $t_m$ is the time label for each time bin, $x_m$ is the number of counts in 
these bins, $N_{ph}$ is the total number of photons in each light curve and C is the average count rate in each time
segment used to construct PSD. The Miyamoto normalization is in units of (rms/mean)$^2$/Hz 
measuring the variability
amplitudes in different energies and can be integrated over different frequency 
range of interest. The top panel of Figure 1 shows the average PSD of the persistent
emission light curve for the  
\XMM observation 
of \thirteen\ computed using the Miyamoto normalization. 
Finally,
for integrated frequency ranges of interest, the frequency dependent spectra can be 
constructed by 
$$ F(E_k,f_m)=C_k \sqrt{P_k(f_j) \Delta f_j} $$ where F is the count rate of the spectrum on the 
frequency $f_m$ in the energy channel $E_k$.

\begin{figure}
\includegraphics[width=8cm,height=8.5cm,angle=270]{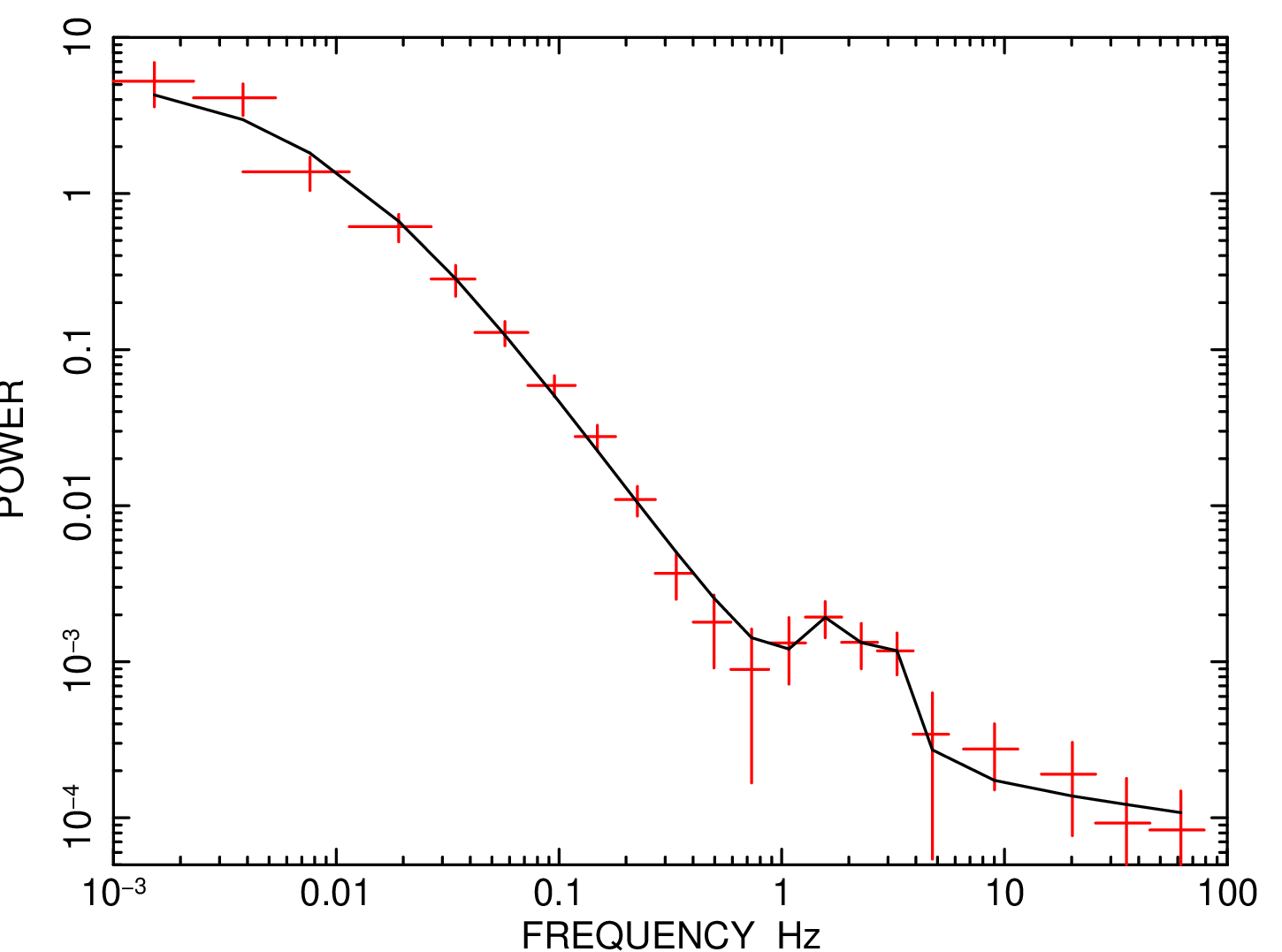}\\

\includegraphics[width=8cm,height=8.5cm,angle=270]{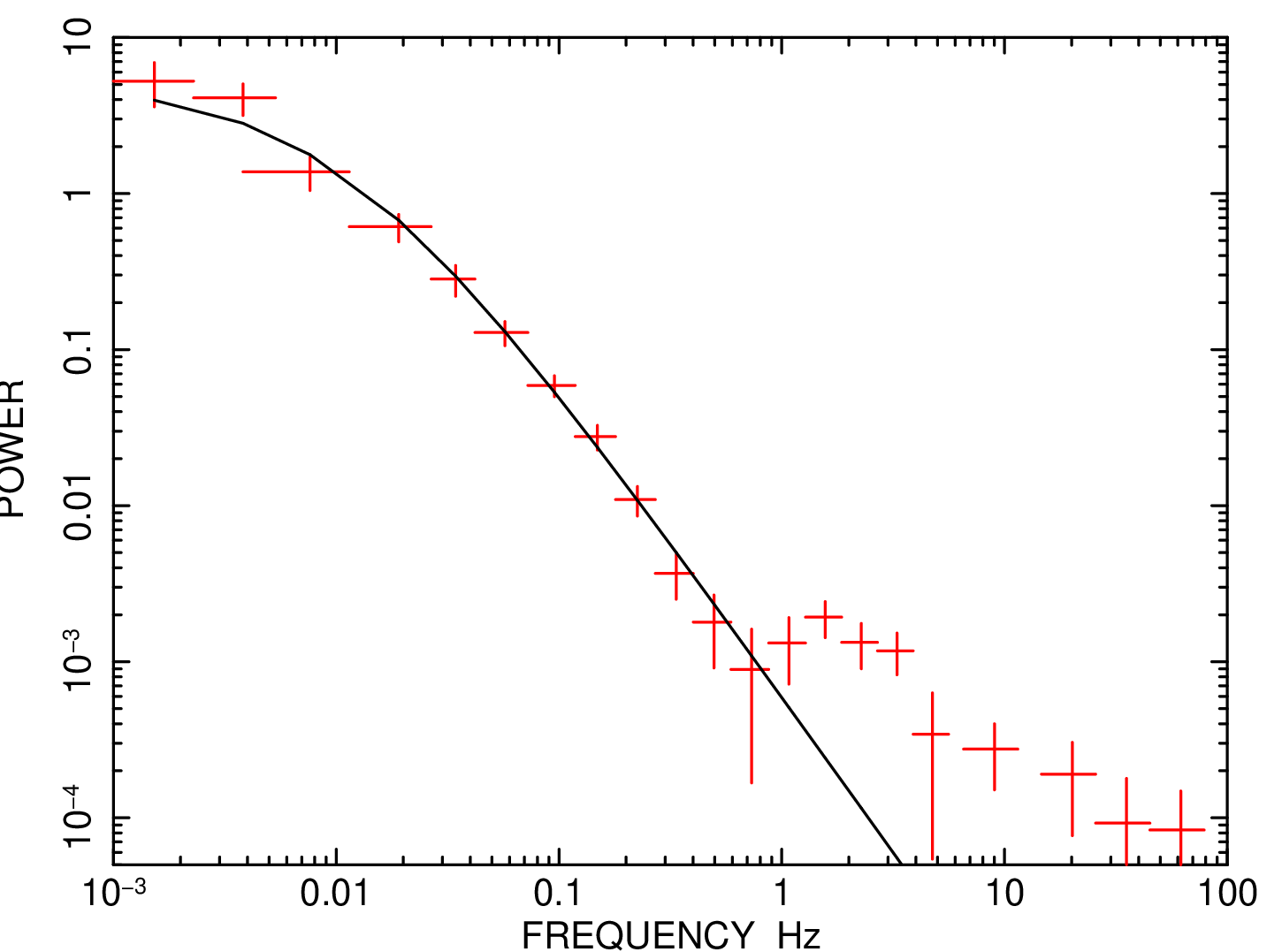}
\caption{The average power spectrum (PSD) of \thirteen\ using a Miyamoto 
normalization in units of (rms/mean)$^2$/Hz. A certain white noise level is subtracted.
The top panel is the PSD of the persistent emission data 
fitted with a composite model of three Lorentzians and a power-law model.
The bottom panel shows the fit to the same PSD with a single Lorentzian
displaying a high frequency excess. }

\end{figure}   

We calculated FRS with the above method in four different frequency bands 
of 0.002-0.04 Hz, 0.07-0.3 Hz, 0.9-6 Hz, and 8-30 Hz. 
The frequency bands were selected according to the simple criteria that there is
good statistics in the frequency range and the FRS produced have comparable
count rates. We also used frequency ranges that complies with the
different components of the PSD. Among these bands,
0.9-6 Hz frequency range (the two narrow Lorentzian components of the PSD as 
discussed below in the next section) includes  the frequency of 1 Hz-QPO 
detected earlier from this system (Jonker et al. 1999).

\subsection{The Average Power Spectrum and the Detected QPOs}

We averaged several power spectra to create a PSD of \thirteen\ using the Miyamoto 
normalization in units of (rms/mean)$^2$/Hz.
Figure 1 shows a fitted PSD with a composite model of three Lorentzians and a power-law following the prescriptions of Belloni, Psaltis, $\&$ van der 
Klis (2002),
yielding a $\chi^2_{v}$\ of 1.0 for 10 degrees of freedom 
(the white noise level is already subtracted). The bottom panel of Figure 1 displays
the same PSD fitted by a single broad Lorentzian producing an unsuccessful fit with
a $\chi^2_{v}$\ value of 2.9 for 17 degrees of freedom.
The very broad Lorentzian models the low frequencies. One of the narrow Lorentzians
have a peak at 1.4$^{+1.0}_{-0.2}$ Hz with a width of about 0.001 Hz. This is similar
to the detected 1 Hz QPO from this system. The other narrow Lorentzian is at
2.8$^{+0.2}_{-0.2}$ with a width of 0.003 Hz. The significance of the
detected Lorentzians can be calculated following (Van der Klis, 1989
; see also Boirin et al. 2000).
We assume that SNR (in $\sigma$) = P$_m$-P$_{ref}$/E$_P$ 
where P$_m$ is the power at Lorentzian peak of interest, P$_{ref}$ is the 
power of the continuum noise
at the same peak frequency, and E$_P$ is the error of the frequency bin at the
Lorentzian peak. We note that the standard deviation of the average of the
powers at the  Lorentzian peak is taken as the error since it is larger
and more representative rather than a theoretical estimation including power
at the peak frequency and square root of the multiplication of the
number  of PSD averaged together and the number of raw frequency bins 
averaged together. 
We find that the significance of the
1.4$^{+1.0}_{-0.2}$ Hz Lorentzian is at 2.8 $\sigma$ 
and the 2.8$^{+0.2}_{-0.2}$  Hz Lorentzian is at 3.1 $\sigma$ confidence level,
respectively. In addition, we calculate an integrated rms 
variability  of 15$\%$  for the
QPO at 1.4 Hz and 11$\%$ for the QPO at 2.8 Hz.
Finally, the continuum of the high 
frequencies is best modeled using a power-law with an index of 
-0.4$^{+0.2}_{-0.2}$.  

\subsection{Results of the FRS Analysis}

The derived spectra (FRS) were fitted with a set of models comprising
a combination of, or solely by, 
a power-law, a blackbody and a Gaussian emission line  models, all absorbed by
neutral matter with column density $N_H$  
(these are POWER, BBODY, GAUSS, and WABS models within {\tt XSPEC}, respectively). 
The fit parameters  are displayed in Table 1. 
For the spectral analysis of the derived FRS, {\tt XSPEC} version 12.5.1 has been used (Arnaud 1996). 
Spectral uncertainties are given at  90$\%$
confidence level ($\Delta$\chisq = 2.71 for a single parameter).
We grouped the FRS spectral energy channels in
groups of 2-3 to improve the statistical quality of the spectra.
During the entire fitting procedure a constant systematic error of 13$\%$ was used (
see sec. 2). The fits were conducted within 0.5-9.0 keV range.
Figure 2 shows the fitted spectra in the four frequency bands. 
The conventional spectral analysis
can be found in Boirin et al. (2005) and Church et al. (2005). 

The results show that the best fitting model in the 0.002-0.04 Hz, and 
0.07-0.3 Hz ranges, is a single blackbody (see Figure 2 top panels)
with a $\chi^2_{v}$\ value of 1.0 (see Table 1; 
The best fit results for 
each low frequency FRS are highlighted in bold-face) . 
The blackbody temperature is in a range 
1.0-1.4 keV (90$\%$ confidence level errors). For the
two low frequency ranges, the power-law (POWER) model or 
a power-law plus a Gaussian emission line (POWER+GAUSS) model fits 
yield $\chi^2_{v}$\ values in excess of 4.0 for very similar degrees of freedom of 10-13.
The fits with blackbody and a power-law (BBODY+POWER) or 
a combination of blackbody, power-law and a Gaussian emission line
(BBODY+POWER+GAUSS) models yield similar $\chi^2_{v}$\ values 
to blackbody (BBODY) fits. The added POWER model gives a power-law index
which is unrealistically
steep. Also, the included GAUSS model gives a normalization for the line of almost
zero (see Table 1).
Thus, we tested the significance of adding the power-law (POWER) and Gaussian
emission line (GAUSS) models
to the blackbody (BBODY) model 
using {\tt FTEST}. The {\tt FTEST} probability yields a value
in a range of 0.40-0.46 for including the POWER and POWER+GAUSS models along with the
blackbody model which is too high. As a result, we find 
that the BBODY+POWER or BBODY+POWER+GAUSS composite-model fits 
are redundant.    

The other two FRS in the 0.9-6 Hz, and  8-30 Hz ranges were
found to be consistent with each other and the fits were
conducted simultaneously
to increase the degrees of freedom. We find that the two FRS
can be fitted best using a combination of a power-law and a Gaussian emission line
(POWER+GAUSS) model with a $\chi^2_{v}$\ of 1.1 for 
8 degrees of freedom compared with any other model or
combination of models (see Table 1; The best fit results for each high frequency
FRS is highlighted in bold-face). An emission line at 6.1-6.6 keV 
(90$\%$ confidence level; best fit value is 6.4 keV) 
has been found necessary to improve the 
power-law fits (see Figure 2 bottom panels) . 
The bottom right-hand
panel in Figure 2 shows the unfolded photon spectra and the line is clearly visible. 
This is the iron  K$_{\alpha}$ fluorescence line indicator
of reflection and/or reprocessing in the system dominating the variability at these 
frequencies. 
The width ($\sigma$) of the iron  K$_{\alpha}$ line is $<$0.5 keV for the 
0.9-6 Hz and 0.7$^{+0.3}_{-0.5}$ keV  for the 8-30 Hz ranges. 
The equivalent width of the line is large
and it is 1.6$^{+0.4}_{-1.2}$ keV  for the  0.9-6 Hz and
1.3$^{+0.7}_{-0.9}$ keV for the 8-30 Hz frequency bands, respectively.  
The fits with a single blackbody model, a single power-law
model or a combination of the two (POWER+BBODY) yield  unacceptable fits (see Table 1).
On the other hand, a blackbody, power-law and Gaussian emission line
(BBODY+POWER+GAUSS) composite model fit gives the same $\chi^2_{v}$\ with the
fitted POWER+GAUSS model (see Table 1). The blackbody temperature is 1.4-2.0 for this
fit and the included blackbody model has a very low normalization in the spectra. 
We tested the
significance of adding the BBODY model to the POWER+GAUSS model fit using  {\tt FTEST}.
The {\tt FTEST} probability yields a value of 0.42 for including the BBODY
model along with the POWER+GAUSS model in the fitting procedure which is too high. 
Thus, a fit with a blackbody plus a power-law and a Gaussian emission line is redundant. 

\begin{figure*}

\centerline{
\includegraphics[width=6cm,height=9cm,angle=270]{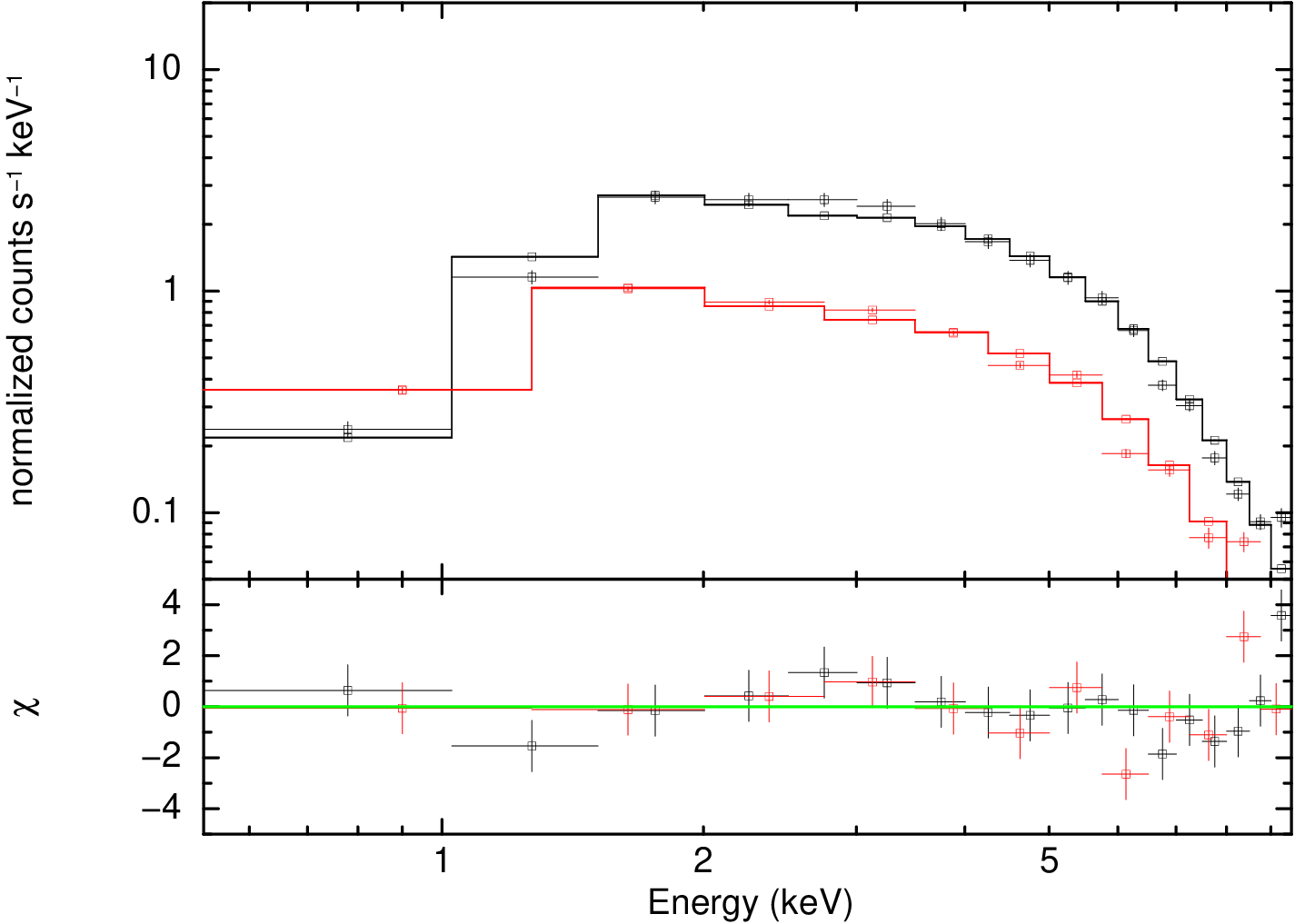} 
\includegraphics[width=6cm,height=8.5cm,angle=270]{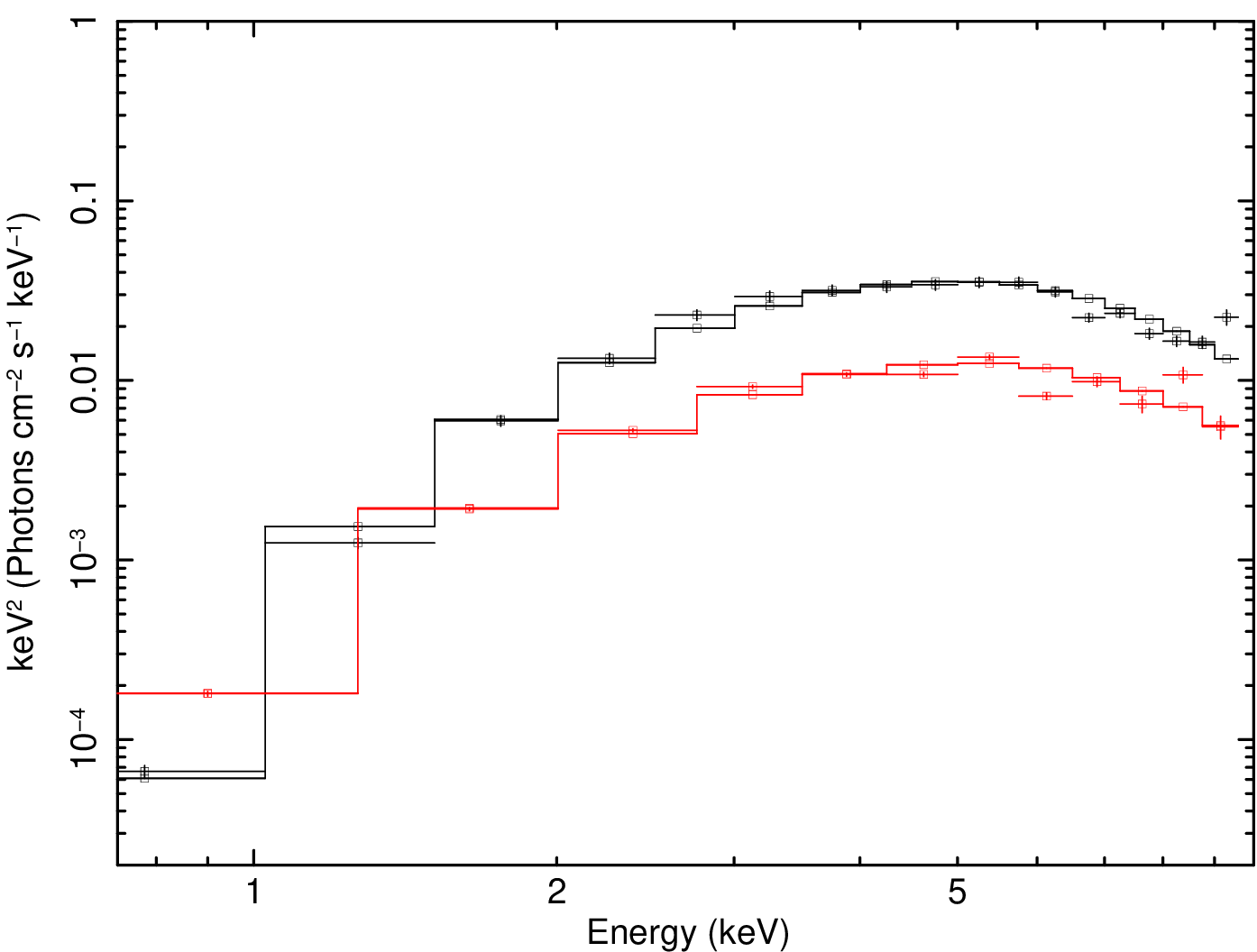}}
\centerline{
\includegraphics[width=6cm,height=9cm,angle=270]{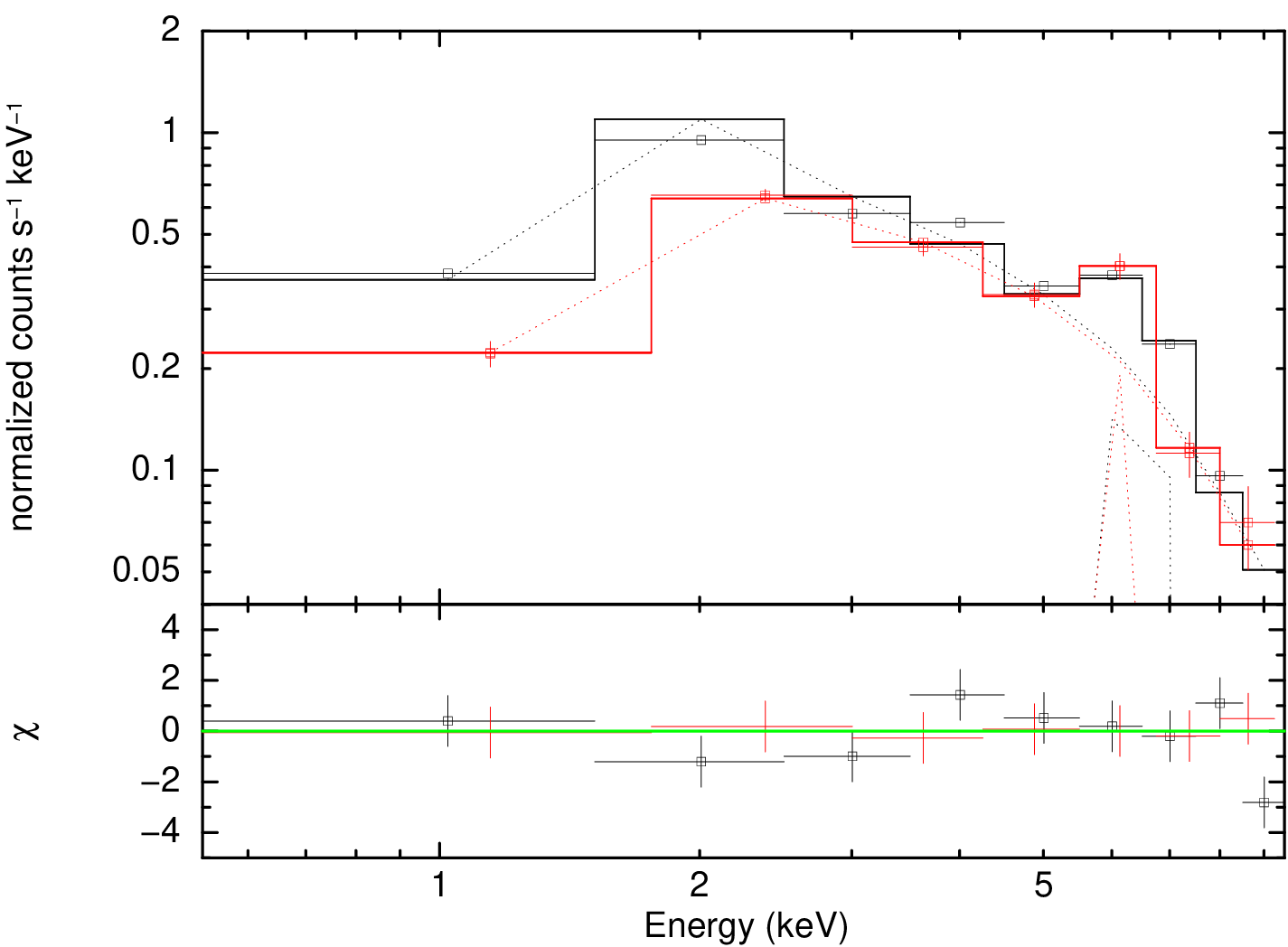} 
\includegraphics[width=6cm,height=8.5cm,angle=270]{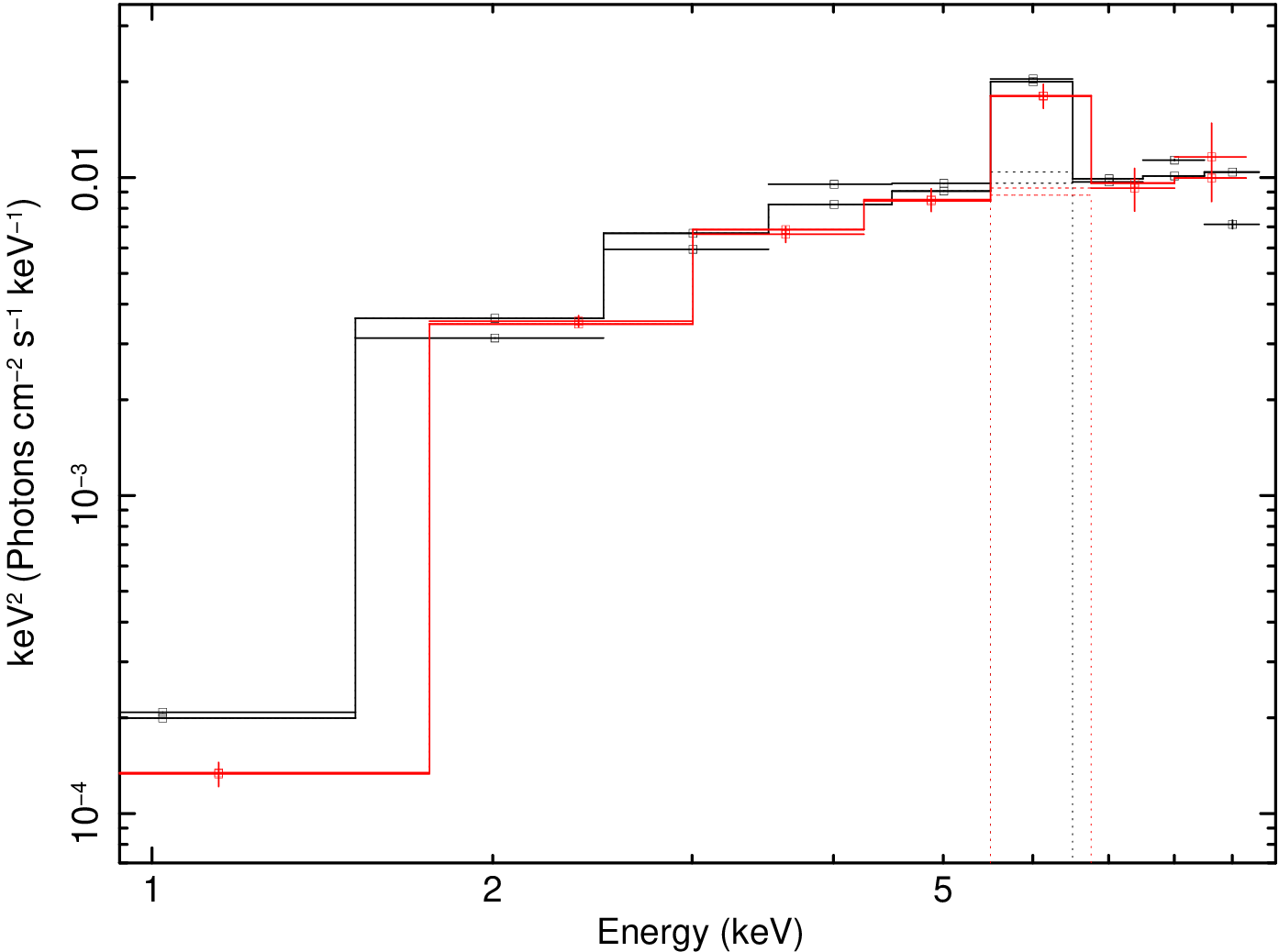}}
\caption{Frequency resolved energy spectra (FRS) of \thirteen\ in four different
frequency bands. The top two panels show the FRS fitted with a 
neural hydrogen absorption model (Galactic)  and a blackbody model 
{\tt WABS$\times$BBODY} (XSPEC models) of emission in the 
0.002-0.04 Hz (the curve in black color), and 0.07-0.3 Hz (the 
curve in red color) frequency bands. 
The top left-hand panel is the count rate spectra and 
the top right-hand panel is the unfolded photon spectra in  {\tt E$^2$f(E) }.
The small panel under the count rate spectra shows the residuals in sigmas.
The two bottom panels show the FRS fitted with a
neural hydrogen absorption model (Galactic) and a power law with a 
Gaussian emission line
model {\tt WABS$\times$(POWER+GAUSS)} (XSPEC models) of emission in the
0.9-6 Hz (the bottom curve in red color) and 8-30 Hz (the top curve in black 
color) frequency bands. The bottom left-hand panel is the count rate spectra and
the bottom right-hand panel is the unfolded photon spectra in {\tt E$^2$f(E)}.
The small panel under the count rate spectra shows the residuals in sigmas.
}
\end{figure*}

\subsubsection{Statistical Significance of the 6.4 keV Fe Line}

We carried out a detailed test of significance of existence of  the
Gaussian emission line using Monte Carlo simulations (e.g., Porquet et al. 2004; Minuitti \& Fabian 2006). 
For our null, hypothesis, we assumed that the spectrum is
simply an absorbed power-law continuum with the same parameters as the absorbed
power-law model fitted to real data. We used the {\tt XSPEC} FAKEIT command to create
1000 fake EPIC pn spectra  corresponding to this model with photon statistics
expected from $\sim$ 40 ksec exposure and grouped the spectra exactly as we have grouped
the real spectra. Next, we fitted each fake spectrum using an absorbed power-law model
with all parameters let to vary to obtain a $\chi^2$ value. We, then added a
line, restricting the line with $\sigma$ between 0-1.0 keV in steps of 0.1 keV and,
constraining the line energy between 6.0-7.5 keV with a step of 0.2 keV  between each
energy value and fitted the fake spectra, this time, with a 
composite model of
a power-law plus a Gaussian line model (GAUSS+POWER). This resulted in 1000 simulations
of $\chi^2$ values from each hypothesis. Then, the two $\chi^2$  values from each 
hypothesis were subtracted. The resulting distribution of  $\Delta\chi^2$  against
cumulative number of occurrences is displayed in Figure 3. The curve is a 
Gaussian centered
on $\Delta\chi^2$$\sim$ 34.2 and varies between 25 and 45. 
The improvement in the $\chi^2$ value as calculated using the 
(POWER and POWER+GAUSS) 
model fits in Table 1 is 33 for 3 degrees of freedom. The probability that 
one gets a detection of a (false) Gaussian line in the fake data
which gives an improvement in $\chi^2$  equal or 
better than that obtained in the real data is measured by integrating 
(counting the number of occurances) Figure 3
for $\Delta\chi^2$$\ge$ 33. This integration yields a probability of 0.002 
implying that the line detection is over 
99$\%$ confidence level
and that the detection in our FRS is not by chance.

\begin{table*}
\caption {Best-fit spectral parameters of FRS for the four different
frequency ranges obtained using a
power-law, a {\tt BBODY} model or the combination of the two 
together with a neutral hydrogen
absorption model {\tt WABS}. Also, an additive Gaussian model is used.
\nh\ is the absorbing column in units of
$10^{22}$~atom~cm$^{-2}$, $kT$
is the blackbody temperature in keV,
$K{\rm _{B}}$ is the normalization of the
blackbody model,
$\Gamma$ is the power law photon index, $K{\rm _{pl}}$ is the model
normalization photon~cm$^{-2}$~s$^{-1}$~keV$^{-1}$ at 1 keV,
$E_0$ is the central energy of the Gaussian model in keV, $\sigma$ is the width
of the Gaussian line in keV, $K{\rm _{G}}$ is the normalization of the Gaussian line
in total photons~cm$^{-2}$~s$^{-1}$~keV$^{-1}$. The best fitting model to the data 
in each frequency range is highlighted in bold-face.
}
\begin{tabular}{llllll}
\hline \hline\noalign{\smallskip} \multicolumn{1}{l}{Model} &
\multicolumn{1}{l}{Parameters} &
\multicolumn{4}{c}{\ Frequency band of the spectra } \\
\noalign{\smallskip\hrule\smallskip}
 &  &  0.002-0.04 Hz & 0.07-0.3 Hz & 0.9-6.0 Hz  & 8.0-31.0 Hz \\
\noalign{\smallskip\hrule\smallskip}
BBODY &  &   &  &   &  \\
\noalign{\smallskip\hrule\smallskip}
 &  N${\rm _H}$ & {\bf 0.4$^{\bf +0.3}_{\bf -0.3}$} & {\bf 0.9$^{\bf +0.3}_{\bf -0.3}$} &  0.9$^{+0.8}_{-0.6}$ & 0.4$^{+0.3}_{-0.3}$  \\
 & $kT_{\rm B}$ &  {\bf 1.3$^{\bf +0.1}_{\bf -0.1}$}   & {\bf 1.2$^{\bf +0.1}_{\bf -0.2}$}  & 1.5$^{+0.4}_{-0.2}$  & 1.5$^{+0.2}_{-0.2}$  \\
 & $K{\rm _{B}}$   & {\bf 0.0003$^{\bf +0.0001}_{\bf -0.0001}$} & {\bf 0.0009$^{\bf +0.0002}_{\bf -0.0001}$}  & 0.0003$^{+0.0004}_{-0.0001}$  & 0.0004$^{+0.0001}_{-0.0001}$  \\
 & $\chi^2_{v}$\ (dof) & {\bf 1.0 (12)} & {\bf 1.0 (13)} & 2.8 (11)  & 2.8 (11) \\
\noalign{\smallskip\hrule\smallskip}
POWER &  &   &  &   &  \\
\noalign{\smallskip\hrule\smallskip}
 & N${\rm _H}$ & 1.2$^{+0.4}_{-0.4}$ & 2.5$^{+2.5}_{-1.0}$ & 2.5$^{+1.0}_{-0.8}$  & 1.5$^{+0.5}_{-0.5}$       \\
 & $\Gamma$  & 1.7$^{+0.6}_{-0.4}$ & 2.4$^{+1.3}_{-0.9}$  & 1.5$^{+0.4}_{-0.4}$  & 1.5$^{+0.4}_{-0.4}$     \\
 & $K{\rm _{pl}}$ & 0.007$^{+0.002}_{-0.001}$  & 0.06$^{+0.01}_{-0.02}$  &  0.005$^{+0.004}_{-0.002}$ &  0.007$^{+0.002}_{-0.004}$ \\
 & $\chi^2_{v}$\ (dof) & 3.8 (12)  & 4.2 (13)  & 3.8 (11)   & 3.8 (11)      \\
\noalign{\smallskip\hrule\smallskip}
POWER+GAUSS &  &   &  &   &  \\
\noalign{\smallskip\hrule\smallskip}
 & N${\rm _H}$ & 1.2$^{+0.4}_{-0.4}$  & 2.2$^{+1.9}_{-0.6}$ & {\bf 2.8$^{\bf +0.6}_{\bf -0.8}$}  & {\bf 1.8$^{\bf +0.4}_{\bf -0.4}$}       \\
 & $\Gamma$  & 1.8$^{+0.5}_{-0.5}$  & 2.2$^{+1.0}_{-0.6}$  & {\bf 1.8$^{\bf +0.3}_{\bf -0.2}$}  & {\bf 1.8$^{\bf +0.3}_{\bf -0.2}$}     \\
 & $K{\rm _{pl}}$ & 0.008$^{+0.001}_{-0.002}$ & 0.04$^{+0.01}_{-0.01}$  &  {\bf 0.007$^{\bf +0.004}_{\bf -0.002}$} &  {\bf 0.007$^{\bf +0.004}_{\bf -0.002}$}  \\
 & $E_0$  & 6.4 (fixed)  & 6.4 (fixed) & {\bf 6.4$^{\bf +0.2}_{\bf -0.3}$}  & {\bf 6.4$^{\bf +0.2}_{\bf -0.3}$}   \\
 & $\sigma$  & 0.2 (fixed) & 0.2 (fixed) & {\bf $\bf <$0.5 } & {\bf 0.7$^{\bf +0.3}_{\bf -0.5}$}   \\
 & $K{\rm _{G}}$  & $<$0.0001 & $<$2.5$\times$10$^{-5}$ & {\bf 0.0003$^{\bf +0.0001}_{\bf -0.0001}$} & {\bf 0.0003$^{\bf +0.0001}_{\bf -0.0001}$}   \\
 & $\chi^2_{v}$\ (dof) & 4.6 (10)   & 4.9 (11)   & {\bf 1.1 (8)}   & {\bf 1.1 (8)}      \\
\noalign{\smallskip\hrule\smallskip}
POWER+BBODY &  &   &  &   &  \\
\noalign{\smallskip\hrule\smallskip}
 & N${\rm _H}$ & 0.7$^{+1.2}_{-0.4}$ & 1.2$^{+0.8}_{-0.2}$   & 2.3$^{+0.5}_{-0.6}$  & 2.3$^{+0.5}_{-0.6}$       \\
 & $\Gamma$  & 5.0$^{+2.8}_{-2.8}$ & 9.9$^{<}_{-6.3}$ & 4.0$^{+2.2}_{-0.5}$  & 4.0$^{+2.2}_{-0.5}$     \\
 & $K{\rm _{pl}}$ & 0.001$^{+0.001}_{-0.0009}$   & 0.001$^{+0.004}_{-0.0009}$  &  0.03$^{+0.02}_{-0.01}$ &  0.03$^{+0.02}_{-0.01}$ \\
 & $kT_{\rm B}$ &  1.2$^{+0.1}_{-0.1}$  & 1.2$^{+0.2}_{-0.2}$  & 1.5$^{+0.06}_{-0.7}$  &  1.5$^{+0.06}_{-0.7}$ \\
 & $K{\rm _{B}}$   & 0.0003$^{+0.0001}_{-0.0001}$ & 0.0009$^{+0.0002}_{-0.0001}$  &  0.0004$^{+0.0004}_{-0.0003}$ & 0.0004$^{+0.0004}_{-0.0003}$  \\
 & $\chi^2_{v}$\ (dof) & 1.0 (10)  & 1.0 (10) &  3.8 (11)  &  3.8 (11)   \\
\noalign{\smallskip\hrule\smallskip}
POWER+BBODY+GAUSS &  &   &  &   &  \\
\noalign{\smallskip\hrule\smallskip}
 & N${\rm _H}$ & 0.7$^{+1.2}_{-0.4}$ & 1.2$^{+0.8}_{-0.2}$ & 3.9$^{+2.0}_{-0.6}$  & 3.9$^{+2.0}_{-0.6}$       \\
 & $\Gamma$  & 4.0$^{+3.4}_{-0.8}$  & $<$9.6 & 2.1$^{+0.3}_{-0.3}$  & 2.1$^{+0.3}_{-0.3}$     \\
 & $K{\rm _{pl}}$ & 0.001$^{+0.003}_{-0.0009}$  &  0.002$^{+0.003}_{-0.001}$   &  0.01$^{+0.02}_{-0.003}$ &  0.01$^{+0.02}_{-0.003}$ \\
 & $kT_{\rm B}$ & 1.2$^{+0.1}_{-0.2}$   & 1.1$^{+0.2}_{-0.1}$  &  1.7$^{+0.3}_{-0.2}$ & 1.6$^{+0.4}_{-0.2}$ \\
 & $K{\rm _{B}}$   & 0.0003$^{+0.0001}_{-0.0001}$ & 0.001$^{+0.001}_{-0.0009}$ & 0.00008$^{+0.00007}_{-0.000002}$  &  0.00009$^{+0.00005}_{-0.00003}$ \\
 & $E_0$  & 6.4 (fixed)  & 6.4 (fixed)  & 6.5$^{+0.2}_{-0.5}$ & 6.5$^{+0.2}_{-0.5}$  \\
 & $\sigma$  & 0.2 (fixed)  & 0.2 (fixed) & $<$0.5 & 0.8$^{+0.3}_{-0.6}$  \\
 & $K{\rm _{G}}$   & $<$0.0003  & $<$2.4$\times$10$^{-5}$ & 0.0003$^{+0.0002}_{-0.0002}$ &  0.0003$^{+0.0002}_{-0.0002}$    \\
 & $\chi^2_{v}$\ (dof) & 1.0 (9)  & 1.0 (10) & 1.1 (6)   & 1.1 (6)      \\
\noalign{\smallskip\hrule}
\end{tabular}

\label{tab:spectra}

\end{table*}

\section{Discussion}

FRS technique is applied to Galactic black hole candidates (Revnivtsev et al. 1999; Gilfanov et al. 2003;
Sobolewska \& Zycki 2006; Reig et al. 2006) to neutron star LMXBs (Gilfanov \& Revnivtsev 2005, 
Revnivtsev \& Gilfanov 2006; Shrader, Reig \& Kazanas 2007) and also to 
active galactic nuclei (Papadakis et al. 2005, 2006). The systematic application of this technique
to Z sources in several different states of the sources yields soft and hard variable thermal components.
The normalization of these components change with accretion rate and thus state of the source.
The softer of the two blackbodies in question is identified with the accretion disk and the harder one with the
boundary layer. Particularly QPOs from such systems at the horizontal and normal branch indicates a particular
hard FRS revealing the expectations that the QPOs originate from the boundary layer.
The temperature of the harder disk component is found to change slowly with luminosity, but the boundary layer
temperature remains constant with frequency confirming that the boundary layer is dominated with
radiation pressure. These components appear mostly in the relatively higher Fourier frequency ranges of 
spectra whereas the disk-blackbody components appear in the lower Fourier frequency ranges. 

A systematic
FRS analysis of an atoll source (4U 1728-34) performed by Shrader et al. (2007) shows that the island 
state of the
source is dominated by a power law or a blackbody (kT = 1.4-1.7) 
plus a power law model that does not vary with Fourier frequency for 0.008-0.8 Hz, 0.8-8 Hz, 
and 8-64 Hz frequency ranges.
They do not detect a resonance iron line around 6.4 keV  in the FRS. 
The higher accretion rate states of the source
at the lower banana and upper banana show only a harder blackbody with a temperature around kT = 2.0-2.2 keV
indicating that at higher luminosities and accretion rates the variable spectral component is coming from the
boundary layer only and is fairly constant in normalization and temperature across frequencies and different 
luminosities.
Gilfanov et al. (2003) study the FRS of the atoll source 4U1608-52 in the frequency ranges corresponding to
the frequency average spectrum, kHz QPOs and a 45 Hz QPO spectra. Their results show that the FRS is 
consistent with two components one of which is the emission from the boundary layer with kT=2.0-2.2 keV or
a CompTT model of Comptonized plasma emission. However, they find that the QPO spectra are mainly consistent with 
2.1-2.4 keV blackbody spectra revealing the radiation pressure dominated boundary layer emission. 

Our results indicate a blackbody model of emission with kT=1.0-1.4 keV for the two lowest 
frequency ranges from 0.002 to 0.3 Hz. This 
is consistent with the findings of  Shrader et al. (2007) in similar frequency ranges 
with this study. The conventional spectral analysis of the \XMM data shows 
the existence of a blackbody model of emission as the softer X-ray component with
kT=0.92-1.2 keV (Boirin et al. 2005; Church et al. 2005) 
which is, then, a variable component in the spectrum of the source.
However, we note that any FRS calculated in this work shows $F_{\rm FRS}$ $<$ 0.12$F_x$
where $F_x$ is the total X-ray flux inferred from conventional spectral analysis. 
The X-ray luminosity of \thirteen\ (5.2$\times$10$^{36}$ erg\ s$^{-1}$; Boirin et al.
2005) corresponds to         
a mass transfer rate of 0.08$\dot{M}_{edd}$ (The distance is 10-20 kpc and $\dot{M}$ could be as high as
0.13$\dot{M}_{edd}$) . We suggest that this soft component originates in the 
boundary layer
and is persistent over the two frequency ranges supported by radiation pressure as expected.
In addition, $INTEGRAL$ (an average spectrum over two years, Balman 2009), 
$RXTE$ (Barnard et al. 2001) and $Suzaku$ (Balucinska-Church et al. 2009) 
spectral analysis indicate similar blackbody temperatures between 1.0-1.7 keV 
(overlapping within 95$\%$ confidence level error range) taken at different epochs and relatively different luminosities
in a range 1-5.2$\times$10$^{36}$ erg\ s$^{-1}$. This supports the stability of 
an existing  boundary layer 
emission (see also sec. 1).  On the other hand, we need to stress that the
appearance of blackbody emission in the low frequencies such as in this case is a typical
feature of the FRS of Z sources (e.g., Gilfanov \& Revnivtsev 2005) and the
blackbody is, then, a disk-blackbody. However, a disk-blackbody with such
temperatures would be atypically high for an Atoll source (like \thirteen)
(e.g., Tarana, Bazzano $\&$ Ubertini 2008; Di Salvo et al. 2000).  

We find a different form of FRS and hardening of the spectra in the two higher frequency ranges.
The existence of a power-law spectra with an iron fluorescence K$_{\alpha}$ line
at 6.4$^{+0.2}_{-0.3}$ keV in the two frequency ranges 0.9-6 Hz and 8-30 Hz  
with the lowest normalization reveals reflection
from a media located in the vicinity of the main source of emission.  In the case for 
reflection from optically thick cold neutral medium, the main reflection features are
a narrow unshifted iron fluorescence K$_{\alpha}$ line at 6.4 keV, an absorption edge at 7.1 keV 
and a reflected
continuum emission peaked at 20-30 keV (Basko et al. 1974; George \& Fabian 1991). 
However, deviations from this 
reflection model (e.g., broad lines and smeared edges) is detected in X-ray binaries indicating 
complicated 
ionization stages in or motion of the reflecting media (e.g., Miller 2007, Cackett et al. 2009a, 2009b, 2008;
Reis, Fabian \& Young 2009; di Salvo et al. 2009; Reynolds et al. 2010).
The complicated modelling of these modified lines show a line centered mainly on 6.4 keV (errors 
in a range from 6.0 to 6.9 keV) with
large broadening of the widths (equivalent widths in a range 50-400 eV) 
including extended red-wings due to general relativistic effects (i.e., redshifts) or
Comptonization in a hot inner corona.  
In the case for \thirteen\ the spectral parameters on Table 1 reveal that
a cold/warm  reflection model with the unshifted iron K${\alpha}$ line is plausible and 
in the 0.9-6 Hz range. However,
we calculate a $\sigma$ $<$ 0.5 keV for the width of the
iron  fluorescence line in the 0.9-6 Hz range and a $\sigma$ of  0.7$^{+0.3}_{-0.5}$ keV
in the 8-30 Hz frequency band at a 90$\%$ confidence level.
The error on the central energy of the line (i.e., 6.1-6.6 keV) and the large width
of the line (particularly in the 8-30 Hz range) allows for shifted and broad resonance Fe line
in the FRS, as well. 
We detect broadening of the line width with increasing frequency as
was detected in the FRS analysis of Cyg X-1 by Revnivtsev et al. (1999). In general, 
the parameters of the line is similar to the modified 6.4 keV Fe fluorescence lines detected 
in other X-ray binaries.

We calculate that the neutral hydrogen absorption, also, increases by a factor of 3-4
in the spectra at the two high frequency ranges.
This strongly suggests that the FRS method may have revealed the location of an absorbing medium 
in the persistent emission. Particularly, the 0.9-6 Hz FRS shows the highest absorption 
due to neutral Hydrogen as modeled in Table 1, but since the FRS have low spectral 
resolution, the true absorption may be of any kind, warm or cold. 

An emission line at 6.6 keV is also detected in the conventional spectral analysis
of the persistent spectrum (Boirin et al. 2005). 
The \XMM analysis reveals a weak broad line with a FWHM of 2.0$^{+0.6}_{-0.4}$ keV. 
As mentioned in the above paragraphs processes like 
relativistic broadening, Compton Scattering, rotational velocity broadening may be the
reason for this large  line broadening. 
The maximum limits  of the line width 
we calculate from the FRS are smaller than the FWHM derived from conventional spectral
analysis. Since the line in the conventional spectrum is weak, it is 
sensitive to the choice of the continuum emission. However,
the large FWHM is a strong indication that it may be a blending between the 
iron  fluorescence lines originating from the different parts of the disk (i.e., inner
and outer disk).  
                 
We believe that the two high frequency band spectra belong to the reflected emission and thus show a 
different spectra in comparison with the rest of the frequency bands that show 
blackbody emission from the disk.  
The conventional spectral analysis of the \XMM data shows
a power-law emission component as the harder X-ray component with
$\Gamma$=1.6-2.0 (Boirin et al. 2005; Church et al. 2005) which is consistent
with our results.

We find that the PSD of the persistent emission shows two
Lorentzian components around 1.4 and 2.8 Hz 
at about 99$\%$ confidence level. We belive this is consistent with
the $\sim$ 1 Hz QPO previously detected from this system  (Jonker et al. 1999).
There has been two suggestions for
the origin of the 1 Hz QPO: 1) quasi-periodic obscuration of the central source 
by a structure in the disk producing relatively energy-independent 
oscillations (Jonker et al. 1999); 2) global disk oscillations (Titarchuck
\& Osherovich 2000).  Jonker et al. (1999) suggest that
the QPO can not be explained with Compton scattering alone since that would result in strong energy
dependence of the QPO which is inconsistent with their findings. 
They suggest that the relatively low energy 
dependence may arise in a model with an intermediate temperature structure containing hot and cold
electrons combining Compton scattering with absorption. Quasi-periodic obscuration
of the central source emission by an opaque medium, an orbiting bump on the surface of 
the disk, may cause the QPO. The QPO exists in the persistent emission, the dips and the
bursts consistently revealing the connection to our work that the source emission 
is incident on
a reflector  carrying its signature. The high inclination angle of the
system also plays an important role.
The high frequency FRS (0.9-30 Hz) derived in this study supports this scenario for the QPOs, 
since FRS is best fitted with a power-law confirming the 
Comptonizing nature and the absorption is then supported by the existence of the
6.4 keV Fe K$_{\alpha}$ emission line and higher N$_H$ parameter values from the fits. 
In the light of the spectral results, the QPOs,  
and the 6.4 keV Fe K$_{\alpha}$ emission line, we suggest
that high frequency FRS  relates directly to reflection/reprocessing and 
scattering in the disk and the
frequency bands used to derive the FRS would yield a finite light crossing time 
of the reflector 
${\tau}_{ref}$=l$_{ref}$/c, also indicating the location of this region within the disk. 
For 
f=1--30 Hz, l$_{ref}$ is c/2$\pi$f yielding a maximum size of 4.7$\times$10$^9$ cm for the 
reflector  and a minimum size of 1.6$\times$10$^8$ cm that would also denote 
a minimum observable inner disk size (the inner disk is at or closer than this limit).  
We suggest that the FRS method applied in this work
have recovered the reflected component in the system that had not been found before by
conventional methods.

\thirteen\ exhibits iron Fe XXV and Fe XXVI absorption lines and the system
is viewed relatively close to edge-on. 
In general, this is true for most of the dipping LMXBs. The conventional
spectral analysis indicates a lack of orbital
phase dependence of features except during dips suggesting that a highly-ionized
absorber can be located in a thin cylindrical geometry around the compact object.
This suggested location of the highly ionized absorber in LMXBs is mainly
in the outer disk (Boirin et al. 2005; Diaz-Trigo et al. 2006). In comparison with
this scenario,
our FRS analysis indicates that there is an absorbing region which is associated with the reflection
zone in the inner regions of the disk at 4.7-0.2$\times$10$^9$ cm. 
However, we note that though we find an absorbing medium in the inner regions of
the disk, we do not rule out another absorbing region (possibly highly ionized)
which can exist somewhere in the outer disk.  
A different approach using dip-ingress timing for all the dipping LMXBs 
(Church $\&$ Balucinska-Church 2004) indicates existence of 
extended accretion disk coronae (ADCs) having radial extend typically 50000 km
or 5-50$\%$ of the accretion disk radius (Balucinska-Church et al. 2009). The
calculated size of an extended ADC  in \thirteen\ is $\sim$2.7$\times$10$^9$ cm
(Church $\&$ Balucinska-Church 2004) with a temperature of about 44 keV.
$INTEGRAL$ results show a temperature of about 196 keV in 4-200 keV range (
Balman 2009) with an estimated location of a suggested ADC at r $<$ 1$\times$10$^9$ cm.
The assumed location of the ADC have some overlap with the 
reflection region size we calculated using our FRS. Then, the 6.4 keV line broadening 
and may be due to Comptonization in an ADC. We note that the size of the reflection 
region is 4.7-0.2$\times$10$^9$ cm as mentioned above and the general relativistic effects
are detected within 40-1$\times$10$^7$ cm (e.g., Papitto et al. 2009; Cackett et al.
 2009b; Reis et al. 2009) and may not explain the
large equivalent widths and extended line widths. 
We note that (using the $INTEGRAL$ results) such a hot ADC may possibly show 
band limited noise effects in harder energies and at relatively higher frequencies
than the ones studied in this work. 

Finally, we stress that we do not recover any absorption features (i.e., lines)
in the four FRS we have calculated in this study which may rule out the existence
of a highly ionized absorber within the inner zones of the accretion disk. However,
if there were narrow weak absorption lines restricted to width $\sigma$ $<$ 0.14 keV
and equivalent widths $<$ 0.04 keV (Boirin et al. 2005), our crude spectral
resolution might have missed them out since we detect a strong Fe fluorescence line
at 6.4 keV.
 
\begin{figure}
\centerline{
\includegraphics[width=8cm,height=8.7cm,angle=270]{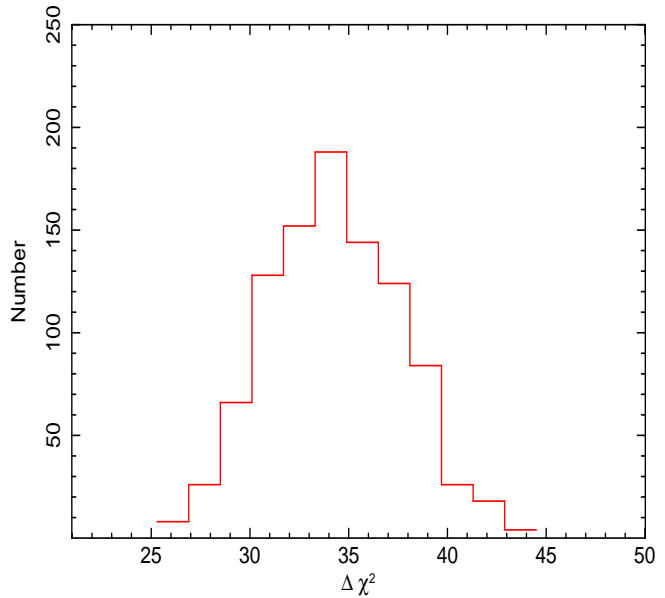}}
\caption{$\Delta\chi^2$ against cumulative number of chance of occurrences for
adding a single Gaussian line in 1000 randomly simulated spectra. The null hypothesis
spectra is a power-law with a neutral hydrogen absorption (Galactic).
The $\Delta\chi^2$ values vary between 25 and 45 for
8 degrees of freedom. This line detection is over 99$\%$ confidence level
in the frequency resolved spectra.
}

\end{figure}

\section{Summary and Conclusions}

We have analyzed the frequency resolved energy spectra (FRS) of the low-mass
X-ray binary dipper \thirteen\ during persistent emission
in four different frequency bands
using an archival \XMM observation.
We recover two different continuum shapes for the four frequency ranges 
we have studied.
We find that the
FRS shows a single blackbody component with kT
in a range 1.0-1.4 keV for the variability in
the frequency ranges 0.002-0.04 Hz, and 0.07-0.3 Hz.
We attribute this component to the accretion disk and possibly emission from
an existing boundary layer supported by radiation pressure.  
On the other hand, we stress that the
emergence of a blackbody component in the
lower frequency ranges and disappearance towards the higher frequencies
suggests that it may also be a disk-blackbody emission given the FRS properties
derived for Z-sources. 
We find a different form of FRS for the frequency
ranges 0.9-6 Hz and 8-30 Hz  which is a power-law model together with a
Gaussian emission line at 6.4$^{+0.2}_{-0.3}$ keV with an equivalent width
of 1.6$^{+0.4}_{-1.2}$ keV and 1.3$^{+0.7}_{-0.9}$ keV for the two frequency
ranges, respectively. The width $\sigma$ of the line is
$<$ 0.5 keV and 0.7$^{+0.3}_{-0.5}$ for the 0.9-6 Hz and 8-30 Hz
ranges. In addition, we recover different neutral hydrogen column 
density for the FRS derived in our work. The $N_{\rm H}$ is highest
in the 0.9-6.0 Hz range with a value of 2.8$^{+0.6}_{-0.8}$ $\times 10^{22}$ cm$^{-2}$
and is only slightly lower for the 8.0-30.0 Hz range, but lower by a factor of 3-4 in the
0.002-0.04 Hz and 0.07-0.3 Hz ranges. So, there is more associated
absorption towards the inner disk and a distinct absorption difference
for the two different continuum detected for the FRS.  
We find that the emission structure within the inner regions of the
disk is such that there is either boundary layer emission and an embedded reflection
region with higher absorption towards the inner regions or there is a
reflection region and an ADC in the inner disk followed by a disk-blackbody
towards the outer disk regions. 
The derived iron fluorescence line shows the existence of
reflection/reprocessing in this system within the inner disk region.
The size of the reflection region is in a range
4.7-0.16$\times 10^9$ cm assuming light travel times.
We suggest that the FRS method applied in this work
have recovered the reflected component in the system that had not been found before by
conventional spectral methods.
We stress that FRS can be used in this manner to find reflection components in other LMXBs
where conventional spectra can not be adequate.
Finally, we find that
the 0.9-6 Hz frequency band shows two QPO peaks at 1.4$^{+1.0}_{-0.2}$ Hz
and 2.8$^{+0.2}_{-0.2}$  Hz at about 2.8-3.1 $\sigma$ confidence level
with integrated rms variability of 15$\%$ and 11$\%$, respectively.
We  relate the origin of the QPOs to the reflection phenomenon. The width of the
6.4 keV line increases with increasing frequency where the location, the QPOs
are produced, is the outer most regions of the reflection zone with the least spread in the
line and the highest $N_{\rm H}$, thus, possibly a cold reflection zone (i.e., region
with large clumps).

\section*{Acknowledgments}
The author thanks an unknown referee for the critical reading
of the manuscript and for very insightful remarks that improved
the manuscript. 
SB, also, thanks to M. Gilfanov for useful comments on the FRS method.
SB acknowledges support from EUFP6 Transfer of knowledge Project
MTKD-CT-2006-042722.
In addition, SB acknowledges support from
T\"UB\.ITAK, The Scientific and Technological Research Council
of Turkey,  through project 108T735.
SB also thanks Tom Marsh and Danny Steeghs for the hospitality
during her visit at the University of Warwick.

\label{lastpage}

\end{document}